\documentclass{article}
\usepackage{amssymb, amsmath}
\usepackage[body={5.5in,8in}]{geometry}
\def\be{\begin{equation}}
\def\ee{\end{equation}}
\def\beq{\begin{eqnarray}}
\def\eeq{\end{eqnarray}}

\def\pd{\partial}

\def\p{\Phi}

\begin{document}
\date{}
\title{Soliton stability in a generalized sine-Gordon potential}
\author{Rub\'en Cordero\footnote{Escuela Superior de F\'{\i}sica y Matem\'aticas, Instituto
Polit\'ecnico Nacional, Ed. 9, Unidad Profesional Adolfo L\'opez
Mateos, 07738 M\'exico D F, Mexico. cordero@esfm.ipn.mx.}
\hspace{.3cm} and Roberto D. Mota\footnote{Unidad Profesional
Interdisciplinaria de Ingenier\'{\i}a y Tecnolog\'{\i}as
Avanzadas, IPN. Av. Instituto Polit\'ecnico Nacional 2580, Col. La
Laguna Ticom\'an, Delegaci\'on Gustavo A. Madero, 07340 M\'exico
D. F. Mexico. mota@esfm.ipn.mx}} \maketitle

\begin{abstract}
We study stability of a generalized sine-Gordon model
with two coupled scalar fields in two dimensions. Topological soliton solutions are found from the first-order equations that solve the equations of motion. The perturbation equations can be cast in terms of a Schr\"odinger-like operators for fluctuations and their spectra are calculated.
\end{abstract}



\section{Introduction}
It is well known that in field theories when a discrete symmetry
is broken domain walls arise. Domain walls have been observed in
condensed matter, for example, in liquid crystals. In the
cosmological context domain walls could appear in phase
transitions in the early universe and have some important
consequences (Vilenkin and Shellard, 1994).

In the domain walls context, there exist classical static
configurations with finite minimum localized energy, see for
instance (Bogomol'nyi, 1976; Prasad, 1975). Several authors have
been interested in coupled scalar fields systems due to their
important physical properties. For example, Peter showed in
(Peter, 1996) that surface current-carrying domain wall arises
when a bosonic charge carrier is coupled to the Higgs field
forming the wall. In (Bazeia, et al., 1997; Riazi, et al., 2001)
was studied linear stability of soliton solutions for a class of
systems of coupled scalar self-interacting fields following  the
standard approach of classical stability.

Besides, as it was extensively shown, Witten's supersymmetric
quantum mechanics (Witten, 1981) is the most recent way to study
solvable as well as perturbative problems (Witten, 1981; Lahiri et
al., 1990; Cooper et al., 1995; Junker, 1996; Cooper et al.,
2001).

Some systems of coupled scalar fields present soliton solutions
and their linear stability have been addressed by means of the
SUSY QM formalism in a $2\times 2$ superpotential realization
(Bazeia et al., 1995; Bazeia and Santos, 1996; de Lima Rodrigues
et al., 1998; Dias at al., 2002). The superpotential satisfies the
Riccati equation associated to the perturbation Hessian and
establishes the stability of the system. This fact is because of
the corresponding supersymmetric operators factorize the
perturbation equation and automatically ensure non-negative equal
perturbation frequencies (Bazeia and Santos, 1996).

The sine-Gordon system has been applied to a wide class of
physical problems like propagation of crystal dislocations,
two-dimensional models of elementary particles, propagation of
splay waves in membranes, Bloch wall motion in magnetic crystals
and magnetic flux in Josephson lines (Rajaraman, 1982). It is well
known that this system in (1+1) dimensions has classical soliton
solutions and their non-dissipative properties could be explained
like a finely-tuned  balance between self-interactions and
dispersion.

A generalization of sine-Gordon system with two coupled real
scalar fields showed an important and rich behavior (Riazi et al.,
2002). The potential of this system consists of a product between
a trigonometric and polynomial functions of the fields. Depending
on the rest energies and the boundary conditions, the spectrum of
solitons could be stable, unstable or meta-stable. The former
classical stability analysis was established by means of numerical
analysis.

Another generalization of coupled sine-Gordon model has been given
as an example of continuously degenerate soliton (Shifman and
Voloshin, 1998). In contrast to the models mention above, this
model involves a highly coupled self-interacting fields with
non-polynomial form. This model has richer structure and dynamics
and deserves further analysis.

In this paper we advocate to the study of linear stability
approximation of the generalized sine-Gordon model proposed in
(Shifman and Voloshin, 1998). In section 2, we give the model
consisting of two couple real scalar fields, we present the first
order differential equations that minimize the energy for the
fields and find a particular solution of them. In section 3, we
show that the stability equations can be analyzed in terms of SUSY
QM formalism and reduce the problem of stability to solve a
Riccati equation associated to the perturbation Hessian. In
section 4, we find the spectra of the fluctuation operator and
explictly show the stability of the soliton solution. Finally, we
give the conclusions of this work.

\section{The Model}

In this paper we consider the generalization of the sine-Gordon
model for two scalar interacting fields given by the following
Lagrangian
\begin{equation}
{\cal L}= \frac{1}{2}(\partial_\mu \Phi)^2
+\frac{1}{2}(\partial_\mu X)^2 - {1\over 2} \cos^2\Phi(1+\alpha
\sin X)^2 -{1\over 2}\cos^2X(1+\alpha \sin \Phi)^2,
\label{potencial}
\end{equation}
where $\alpha$ is a dimensional parameter, and all other
dimensional parameters are set equal to unity. For $0\leq
\Phi,X\leq 2\pi$, the potential in (\ref{potencial}) has three
minima at $\Phi=X=\pi/2$; $\Phi=\pi/2$, $X=3\pi/2$ and
$\Phi=X=3\pi/2$, one maximum at $\Phi=0$, $X=\pi/2$, and three
saddle points at $\Phi=\pi/2$, $X=\pi$; $\Phi=\pi/2$, $X=0$ and
$\Phi=3\pi/2$, $X=\pi/2$.

 The equations of motion for the model
(\ref{potencial}) are the usual ones: \be \square{\Phi} +
\frac{\pd}{\pd \p}V =0, \hspace{2cm} \Box X + \frac{\pd}{\pd X}V
=0 \ee which become for a static configurations \beq \p'' &=& -
\cos \p \sin \p (1+ \alpha \sin X)^2 +
\alpha \cos ^2 X (1+ \alpha \sin X)\cos \p  \\
X'' &=& - \cos X \sin X (1+ \alpha \sin \p)^2 +
\alpha \cos ^2 \p (1+ \alpha \sin \p)\cos X,
\eeq
where primes means derivatives with respect to space variable.

The form of the
energy of the system can be written as
\begin{equation}
E_s=\int_{-\infty}^{\infty}\left[\left(\frac{d\Phi}{dz}-W_{\Phi}\right)^2+
\left(\frac{dX}{dz}-W_{X}\right)^2\right]dz+
{\Big\vert}\int_{-\infty}^{\infty} {\partial\over \partial z}W[\Phi(z),X(z)]dz
{\Big\vert}
\end{equation}
where $W[\Phi(z),X(z)]$ is the corresponding superpotential of
(\ref{potencial}),  which turns out to be
\begin{equation}
W=-\sin\Phi-\sin X-\alpha(\sin\Phi)(\sin X).
\label{superpotencial}
\end{equation}
In  (Shifman and Voloshin, 1998) this superpotential was referred
as a generalization of the sine-Gordon model. It is periodic in
both $\Phi$ and $X$; for $\alpha =0$ it describes two decoupled
fields, representing each of them a supergeneralization of the
sine-Gordon model. If $\alpha\neq 0$ the fields $\Phi$ and $X$
start interacting with each other. Inside the periodicity  domain
$0\leq \Phi,X\leq 2\pi$, $-W$ has one maximum at $\Phi=X=\pi/2$,
one minimum at $\Phi=X=3\pi/2$ and two saddle points at
$\Phi=\pi/2$, $X=3\pi/2$ and $\Phi=3\pi/2$, $\pi/2$ at least for
small values of $\alpha$.

The lower bound for the energy is achieved if $\Phi$  and $X$ satisfy
\begin{eqnarray}
\Phi^{\prime}=-\cos{\Phi}(1+\alpha\sin{X})\nonumber\\
X^{\prime}=-\cos{X}(1+\alpha\sin{\Phi}). \label{monster}
\end{eqnarray}
For the case $X=\pi/2$, we have
\begin{equation}
{d\Phi\over dz}=-\cos{\Phi}(1+\alpha),
\end{equation}
whose solution is
\begin{equation}
\Phi=-\tan^{-1}\left({c^2 e^{z(1+\alpha)}-e^{-z(1+\alpha)}\over 2}\right).
\label{solucion}
\end{equation}
Other possible solution of equations (\ref{monster}) is obtained
for $X=3\pi/2$ and the solution is obtained from the former
equation by substituting $\alpha$ by  $-\alpha$. Interchanging the
fields $X$ and $\Phi$ in the last equations we get the solution
for $\Phi = \pi/2$ or $\Phi=3\pi/2$.

We attempted to find the general solutions of the coupled
equations (\ref{monster}) by the trial orbit method of Rajaraman
(Rajaraman, 1977), however, because of the difficulty of the
system we were unable to find them.

\section{Stability Equations and SUSY QM}

We are interested in determining the classical stability of this
system under small fluctuations around a static configuration. In
order to investigate the linear stability of the interacting
fields we proceed in the usual way by considering small
perturbations around the static scalar fields
\beq
\Phi(z,t) &=&
\Phi(z) + \eta(z,t) \\
X(z,t) &=& X(z) + \xi(z,t). \eeq The stability equations can be
written in a Schr\"odinger-like equation
\be S_l \Psi_n =
{\omega_n}^2{\Psi}_n
\ee
where $n=0,1,2..$. The differential operator
$S_l$ is given by

\begin{equation}
{S}_{l}=\begin{pmatrix} -\frac{d^2}{dz^2} + \frac{\pd^2}{\pd
\p^2}V& \frac{\pd^2}{\pd \p \pd X}V \\
          \frac{\pd^2}{\pd \p \pd X}V&  -\frac{d^2}{dz^2} +
          \frac{\pd^2}{\pd X^2}V
\end{pmatrix}_{|\p =\p(z), X=X(z)}
\equiv  -\frac{d^2}{dz^2} {\mathbf I}_{2\times 2} + {\mathbf
V}_{PH}
\end{equation}
and the two components wave functions are
\be
\Psi_{n}=\begin{pmatrix} \p_n (z) \\
          X_n(z)  \\
\end{pmatrix},
\ee
where we have expanded the fluctuations
$\alpha(z,t)$ and $\beta(z,t)$ in terms of normal modes \beq
\eta(z,t) &=& \sum_{n}a_n \eta_n (z)e^{i\omega_nt} \\
\xi(z,t) &=& \sum_{n}b_n \xi_n (z)e^{i{\omega}_nt}.
\eeq
Notice that in the case when the differential operator $S_l$ is diagonal
the perturbation fields could be expanded in terms of different frequencies.

The SUSY QM approach to linear stability consists in realizing a $2\times
2$-matrix superpotential, which is obtained by solving the Riccati
equation associated to the perturbation Hessian
${\bf V}_{PH}$
 \begin{equation}
{\mathbf W}^2+{\mathbf W}^{\prime}={\mathbf V}_{PH}.
\label{eq:ric}
\end{equation}
The existence of ${\mathbf W}$ that satisfies this equation
ensures the existence of the first order self-adjoint differential
operators \be {\cal D}^{\pm} = \pm {\mathbf I}\frac{d}{dz} +
{\mathbf W}(z) \ee that factorize the operator $S_l= {\cal D}^+
{\cal D}^- $. This fact implies the stability for equal
fluctuation frequencies, since $0 \leq |{\cal D}^- \Psi_n|^2 =
({\cal D}^- \Psi_n)^{\dag}({\cal D}^- \Psi_n)= \langle {\cal D}^+
{\cal D}^- \rangle = \langle S_l \rangle =  \omega_n ^2$.

For our case the matrix elements of ${\mathbf V}_{PH}$ are given by
\beq
({\mathbf V}_{PH})_{11} &=& -(\cos^2 \p -\sin^2 \p)(1+ \alpha \sin X)^2 +
\alpha^2 \cos^2 X \cos^2 \p \\ \nonumber
&-& \alpha \cos^2 X(1 + \alpha\sin \p)\sin \p  \\ \nonumber
({\mathbf V}_{PH})_{22} &=& -(\cos^2 X -\sin^2 X)(1+ \alpha \sin \p)^2 +
\alpha^2 \cos^2 \p \cos^2 X \\ \nonumber
&-& \alpha \cos^2 \p(1 + \alpha\sin X)\sin X  \\
({\mathbf V}_{PH})_{12} &=&({\mathbf V}_{PH})_{21} = -2\alpha\cos \p \sin \p(1 +
\alpha \sin X)\cos X \\ \nonumber
&-& 2\alpha \cos X \sin X( 1 + \alpha \sin \p)\cos \p,
\eeq
The solution of Riccati equation (\ref{eq:ric}) for configurations satisfying equations
(\ref{monster})  is
\begin{equation}
{\mathbf W}_{min}=\begin{pmatrix}(1+\alpha\sin X)\sin{\Phi} &
-\alpha\cos \p \cos X \\
             -\alpha\cos \p \cos X           &(1+\alpha\sin \p)\sin{X} \\
\end{pmatrix}.
\end{equation}
For the sector $X=\pi/2$, the fluctuation potential term becomes
\begin{equation}
{\mathbf
V}_{min}=\begin{pmatrix}-(1+\alpha)^2(\cos^2{\Phi}-\sin^2{\Phi}) &
0 \\
                     0     & (1+ \alpha \sin \p)^2 - \alpha \cos^2 \p (1 + \alpha)\\
\end{pmatrix},
\end{equation}
so, the corresponding superpotential is
 \begin{equation}
{\mathbf W}_{min}=\begin{pmatrix}(1+\alpha)\sin{\Phi} & 0 \\
                     0  & (1 + \alpha\sin{\Phi})\\
\end{pmatrix}.
\end{equation}

We point out the existence of another self-adjoint and
non-negative second-order differential operator
$S_l^{\prime}={\cal D}^- {\cal D}^+ $ which plays the role of the
supersymmetric partner operator of $S_l$ in SUSY QM. The operators
$S_l$ and $S_l^{\prime}$ have the same energy spectrum except for
the ground state.

\section{Spectrum of the Second Order Fluctuation Operator}

The study of stability for the general case is very difficult.
However, in order to have analytical results in the case of
$X=\pi/2$ (the results we are going to obtain are automatically
true for $\Phi=\pi/2$), we take the particular case of $c=1$ in
equation (\ref{solucion}) {\it i. e. } $\tan{\Phi}=\sinh
{z(1+\alpha)}$. Since the differential operator $S_l$ is diagonal
we could have different fluctuation frequencies that can be
determined from the perturbation equations
\begin{equation}
-{d^2\eta_n\over dz^2}-(1+\alpha)\left(2
\mbox{sech}^2{z(1+\alpha)}-1\right)\eta_n= \omega^2_n\eta_n
\label{etaeqn}
\end{equation}
and
\begin{equation}
-{d^2\xi_n\over dz^2}+\left(1+\alpha^2-2\alpha\tanh{z(1+\alpha)} -
\alpha(1 +
2\alpha)\mbox{sech}^2{z(1+\alpha)}\right)\xi_n=\omega^{2}_n\xi_n.
\label{xieqn}
\end{equation}
Performing the variable change $y=z(1+\alpha)$, equation
(\ref{etaeqn}) transforms to the  Rosen-Morse equation (Morse and
Feshbach, 1953). We find that the fluctuation frequencies are
\begin{equation}
\omega_n^2=(1+\alpha)^2\left(1-(1-n)^2\right)^2.
\end{equation}
However the bound states exist only for $n<1$ (Morse and Feshbach,
1953). Thus, the ground state
$\eta_0=(1+\alpha)\mbox{sech}z(1+\alpha)$ with eigenvalue
$\omega_0=0$ is stable.

By means of the same variable change the equation (\ref{xieqn})
can be cast as a Rosen-Morse equation whose eigenvalues are
\begin{eqnarray}
\omega_n^2&=& 1+\alpha^2-(1+\alpha)^2\left[\left( \frac{3\alpha + 1}{2(1+\alpha)} -
(n + 1/2)\right)^2 \right] \\ \nonumber
&-&  \frac{4\alpha}{(1+\alpha)^2\left( \frac{3\alpha + 1}{2(1+\alpha)} - (n + 1/2)\right)^2}
\end{eqnarray}
which are the frequencies for possible bound states. However, for
this case we have no bound states because $n$ must be less than
zero for both $\alpha>0$ and $\alpha<0$ (Morse and Feshbach,
1953).

This means that the solution configurations are stable under small perturbations around
$X=\pi/2$, $\tan{\Phi}=-\sinh {z(1+\alpha)}$ ( the same is true for $\Phi=\pi/2$,
$\tan{X}=-\sinh {z(1+\alpha)}$).

\section*{Conclusions}

We have applied the SUSY QM formalism to study the linear
stability of the Shifman generalization of the sine-Gordon model.
We have shown that stability for soliton configurations is ensured
by solving the Riccati equation for the $2\times2$ superpotential
associated to the non-diagonal perturbation Hessian. The spectrum
of the second order fluctuation operator for the general case is
very difficult to find it. Thus, we have got the fluctuation
spectrum for the particular case $X=\pi/2$ (or $\Phi=\pi/2$), and
we have found analytical solutions and explicitly found that the
system is stable. We notice that equations (\ref{etaeqn}) and
(\ref{xieqn}) can be reduced to a Rosen-Morse equation. On another
hand, the Rosen-Morse equation have been studied from the shape
invariance approach of SUSY QM (Dutt et al., 1988). Thus, each one
of the equations (\ref{etaeqn}) and (\ref{xieqn}) has a scalar
superpotential. Therefore, we have given a complete treatment of
linear stability for the generalized sine-Gordon superpotential
from the point of view of SUSY QM.

\section*{Acknowledgments}

R D Mota would like to thank the Departamento de Matem\'aticas del
Centro de Investigaci\'on  y Estudios Avanzados del IPN where he
was a visitor during the preparation of this work. This work was
partially supported by SNI-M\'exico, CONACYT grant CO1-41639,
COFAA-IPN, EDI-IPN, and CGPI project number 20030642.

\section*{References}
\noindent Bazeia, D., dos Santos, M. J., and Ribeiro, R. F.
(1995). {\it Physics Letters A}, {\bf 208}, 84.

\noindent Bazeia, D., Nascimento, J. R. S., Ribeiro, R. F. and
Toledo, D. (1997).
 {\it Journal of Physics A: Math. Gen.} {\bf 30}, 8157.

\noindent Bazeia, D. and  Santos, M. M. (1996). {\it Physics
Letters A} {\bf 217}, 28.

\noindent Bogomol'nyi, E. B. (1976). {\it Soviet Journal of
Nuclear Physics} {\bf 24}, 449.

\noindent Cooper, F., Khare, A. and Sukhatme, U. (1995). {\it
Physics Reports}\ {\bf 251}, 267.

\noindent Cooper, F., Khare, A. and Sukhatme, U. (2001). {\it
Supersymmetry in Quantum Mechanics}, (World Scientific,
Singapore).

\noindent de Lima Rodrigues, R., Da Silva Filho P. V. and Vaidya,
A. N. (1998). {\it Physical Review D} {\bf 58}, 125023.

\noindent Dias, G. S., Graca E. L. and de Lima Rodrigues, R.
(2002). {\it hep-th/0205195}

\noindent Dutt, R., Khare, A.  and Sukhatme U. P. (1998). {\it
American Journal of Physics} {\bf 13}, 974.

\noindent Junker, G. (1996). {\it Supersymmetric Methods in
Quantum and Statistical Physics}, (Springer-Verlag, Berlin).

\noindent Lahiri, A.  Roy, P. and Bagchi, B. (1990). {\it
International Journal of Modern Physics}\ A {\bf 5}, 1383.

\noindent Morse, P. M.  and Feshbach, H. (1953). {\it Methods of
Mathematical Physics, Vol II} (McGraw-Hill, New York, USA).

\noindent Peter, P. (1996). {\it Journal of Physics A: Math. Gen.
} {\bf 29}, 5125.

\noindent Prasad M. K. and Sommerfield, C. H. (1975). {\it
Physical Review Letters} {\bf 35},760.

\noindent Rajaraman,  R. (1979). {\it Physical Review Letters}
{\bf 42}, 200.

\noindent Rajaraman, R. (1982). {\it Solitons and Instantons},
(North-Holland, Amsterdam).

\noindent Riazi, M. N., Golshan, M. N. and Mansuri, K. (2001).
{\it Int. J. Theor. Phys. Group. Theor. Non. Op. }; {\bf 7} No. 3,
91.

\noindent Riazi, N., Azizi, A. and Zebarjad, S. M. (2002). {\it
Physical Review D} {\bf 66}, 065003.

\noindent Shifman, M. A.  and Voloshin, M. B. (1998). {\it
Physical Review D} {\bf 57}, 2590.

\noindent Vilenkin, A. and Shellard, E. P. S. (1994). {\it Cosmic
Strings and Other Topological Defects} (Cambridge University
Press, Cambridge)

\noindent Witten, E. (1981). {\it Nuclear Physics }{\bf B185},
513.

\end{document}